\documentclass{article}
\usepackage{spconf,amsmath,graphicx,amssymb,multicol}

\title{On decoder-only architecture for speech-to-text and large language model integration}
\name{\begin{tabular}{c}
    Jian Wu, 
    Yashesh Gaur, 
    Zhuo Chen, 
    Long Zhou, 
    Yimeng Zhu, \\
    Tianrui Wang, 
    Jinyu Li, 
    Shujie Liu, 
    Bo Ren, 
    Linquan Liu, 
    Yu Wu
    \end{tabular}}
\address{Microsoft, One Microsoft Way, Redmond, USA}
\begin{document}
%
\maketitle
\begin{abstract}

Large language models (LLMs) have achieved remarkable success in the field of natural language processing, enabling better human-computer interaction using natural language. However, the seamless integration of speech signals into LLMs has not been explored well. The ``decoder-only” architecture has also not been well studied for speech processing tasks. 
In this research, we introduce Speech-LLaMA, a novel approach that effectively incorporates acoustic information into text-based large language models. 
Our method leverages Connectionist Temporal Classification and a simple audio encoder to map the compressed acoustic features to the continuous semantic space of the LLM. 
In addition, we further probe the decoder-only architecture for speech-to-text tasks by training a smaller scale randomly initialized speech-LLaMA model from speech-text paired data alone. 
We conduct experiments on multilingual speech-to-text translation tasks and demonstrate a significant improvement over strong baselines, highlighting the potential advantages of decoder-only models for speech-to-text conversion.

\end{abstract}
\begin{keywords}
decoder-only, LLaMA, LoRA, speech translation
\end{keywords}
\section{Introduction}
\label{sec:intro}
In recent times, the large language models (LLMs) have showcased remarkable achievements across various natural language benchmarks, encompassing question answering, machine translation, language understanding and more \cite{chatgpt,gpt4,brown2020language,chowdhery2022palm,touvron2023llama}. By employing a Transformer-based architecture \cite{vaswani2017attention} and training to anticipate forthcoming tokens within a sequence, this language model excels in contextual learning abilities. Not only does this significantly enhance its modeling prowess, but more importantly, it enables seamless user interaction that effectively connects cutting-edge research with real-world applications.

As speech represents the most innate and instinctive mode of human communication, integrating speech and LLMs will further boost the user experience of human-machine interaction. 
Based on this intuition, several attempts in combining speech signals and large language models were carried out \cite{huang2023audiogpt,shen2023hugginggpt,chen2023x,deshmukh2023pengi}.  Among them, the cascaded approach is the most straightforward solution. In these systems, the speech signal is firstly transformed into word tokens through existing automatic speech recognition (ASR) \cite{E2EOverview} models, and LLM processes the recognized words for downstream tasks. 
Later, inspired by the integration of image information to LLMs \cite{zhu2023minigpt,liu2023visual,gao2023llama,alayrac2022flamingo}, researchers also explored the deep combination of speech signals \cite{chen2023x,deshmukh2023pengi,wang2023viola,zhang2023speechgpt,nachmani2023lms,rubenstein2023audiopalm}.  In \cite{wang2023viola}, the authors proposed to jointly model the speech and text tasks through a unified decoder only network. Similarly, in \cite{rubenstein2023audiopalm}, the authors proposed to optimize the audio token conversion module together with a off-the-shelf LLM. Instead of word pieces, discrete tokens of speech representation from a self-supervised model are used in \cite{zhang2023speechgpt}.

While there have been promising outcomes, several crucial challenges regarding the integration of speech and LLMs still require further exploration. 
Initially, aligning the two modalities (speech and text) using a pretrained LLM poses challenges due to the typically longer sequence length of speech signals compared to text sequences.
Moreover, given the costly nature of training LLMs, finding ways to minimize the overall integration cost while maintaining exceptional performance continues to be a challenging task.
More importantly, considering the remarkable success of the LLMs, it is crucial to explore the untapped potential of using a decoder-only model~\cite{brown2020language,wang2023viola,wang2023neural, zhang2023speak, fu2023decoder} as the backbone network architecture for speech to text processing.

In this study, we aim to tackle the aforementioned challenges by exploring an efficient end-to-end integration of speech and language models. Our approach involves designing a simple yet effective architecture where a large language model that operates on text also incorporates acoustic embeddings. 
This integration enables the LM to condition its transcription or translation of the acoustic information. 
More specifically, our proposed method utilizes a pre-existing LLM and incorporates a acoustic feature compressor and an acoustic encoder introducing only a small number of free parameters. 
Diverging from previous approaches that convert speech into discretized tokens, our model directly maps the continuous representation of speech into the semantic space defined by the LM. 
During the processing stage, the speech feature is initially compressed by the acoustic compressor to reduce the sequence length. Subsequently, the acoustic encoder transforms the compressed speech signal into continuous vectors in the same semantic space of the text that can be consumed by the LLM. The final output is generated through the decoding process of the LLM.

We thoroughly investigate various practical aspects of our proposed model, such as selecting the appropriate acoustic compressor, attention mask, and fine-tuning methods. Additionally, we apply the proposed model to the task of translating speech in 13 different languages into English (EN) text and compare its performance against a strong baseline on CoVoST dataset. Finally, we demonstrate that the decoder-only model, even trained from scratch using only speech-text paired data, exhibits significant potential and several advantages over the commonly employed encoder-decoder architecture in speech processing. In this work, our contribution can be summarized as follows:
\begin{itemize}
    \item We introduce an efficient end-to-end integration method called Speech-LLaMA, which effectively integrates existing text-based large language models with speech processing. We have achieved substantial improvements in translation performance compared to strong baselines on various speech translation (ST) tasks. 
    \item We investigate various practical aspects of the proposed speech-LLM integrations that are crucial for enhancing performance. These aspects include acoustic compression of the acoustic feature, attention mask selection, and fine-tuning strategy.
    \item On large, diverse and real-world data, we show that the decoder-only architecture can be as competitive as the encoder-decoder architecture for speech-to-text tasks. We show that decoder-only to also be more parameter efficient.
\end{itemize}






\section{Related work}
\label{sec:format}

Our model aims at integrating speech signals into large language models, as well as relates to Connectionist Temporal Classification (CTC) feature length compression and low-rank adaptation (LoRA). We discuss these topics in the following.

\subsection{Large language models}
LLMs are generally pre-trained on vast amounts of textual data that span a wide variety of domains and languages.  They usually consist of a stack of transformer layers, following an auto-regressive decoder-only architecture, where each output token is used as the input to predict the next step token. 
In this work, we select LLaMA-7B \cite{touvron2023llama} as the backbone LLM to build the proposed method. LLaMA-7B model consists of 32 Transformer encoder layers with 32 heads and 4096 attention dimension. The tokenizer from the LLaMA work has a vocabulary size of 32,000 which covers a group of languages.

\subsection{CTC compressor}
Connectionist Temporal Classification (CTC) compressor \cite{gaido2021ctc} was proposed to reduce the sequence length via removing the redundant information in the features. It was applied in speech translation task and was shown to yield better memory consumption and performance. The method adds a linear CTC branch in a middle layer of the encoder which is jointly optimized with the main cross-entropy criteria . The hidden representations of the CTC branch are then compressed according to the distributions of the CTC posteriors and are passed to the succeeding layers. The author investigated a few variations within this method of sequence length compression. They found that averaging the consecutive hidden representations (corresponding to consecutive CTC predictions belonging to the same class) gives the best performance.

\subsection{LoRA}
Low-Rank Adaptation (LoRA) \cite{hu2021lora} is a commonly used technology to adapt the large models for new datasets or tasks. It introduces a small amount of free parameters to each Transformer layer of the source large model, while freezing all the original model parameters. 
Specifically, for each weight matrix $W \in \mathbb{R}^{d \times k}$ in a Transformer layer, 2 new  matrices $W_a \in \mathbb{R}^{d \times r}$ and $W_b \in \mathbb{R}^{r \times k}$ are introduced such that $r \ll \min\{d, k\}$. For each matrix multiplication during training, the input $x$ is firstly multiplied with both original weight $W$ and its introduced low-rank approximation $W_a$, $W_b$, then the two outputs are summed to form the output for later computation. Only $W_a$ and $W_b$ are updated during fine-tuning while $W$ keeps frozen, thus significantly reducing the memory footprint during training. 
%


\begin{figure}[!tbp]
\centering
\includegraphics[width=0.48 \textwidth]{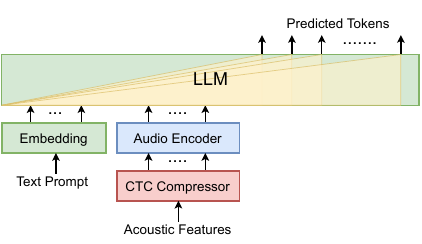}
\caption{High-level architecture of our proposed approach with LLM. The green blocks indicate the part of the LLM. In this work, we only learn parameters in the ``Audio Encoder", keeping everything else frozen.}
\label{fig:res}
\end{figure}

\section{Our approach}
\label{sec:approach}

In this work, we design an architecture named Speech-LLaMA where a text-LLM can also accept acoustic embedding as well as text as conditional prompts for text generation. By converting the speech input to a sequence of acoustic embeddings within the same space of the text embeddings, in the aspect of both length and semantics, the pre-trained text LLM can leverage its in-context learning capacity to absorb the speech signal and output corresponding text for speech translation task. 

Overall, given the text prompt $\mathbf{p}$ and audio signals $\mathbf{x}$, the generation of the corresponding text sequence $\mathbf{y} = \{y_0, y_1, \cdots, y_{N - 1}\}$ with a text-LLM is formulated as:
\begin{equation}
p(\mathbf{y}|\mathbf{p}, \mathbf{x}; \Theta_\text{LLM}) = \prod_{n=0}^{N - 1} p(y_n|\mathbf{y}_{<n}, \mathbf{p}, \mathbf{x}; \Theta_\text{LLM})
\end{equation}
where $\mathbf{y}_{<n}$ indicates the generated text sequence before $y_n$.

\textbf{Overview}
Our proposed neural model consists of three distinct parts: a pre-trained text neural LLM, an audio encoder and a CTC compressor, as shown in Figure 1. The text-LLM in our case is a LLaMA-7B \cite{gao2023llama} but this method can be generalized to LLMs of any scale.  The CTC compressor reduces the sequence length of the input speech filter-bank to match the length of the text, and the audio encoder transforms the compressed speech signal into continuous vectors in the LLM's semantic space. 


\textbf{CTC compressor}
Different from the prior work that trained the CTC compressor jointly with the main task \cite{gaido2021ctc}, our CTC compressor is a pre-trained module, aiming to match the audio and the text duration to the same scale by selecting the representative frames from the audio signal. In this work, we explore two ways to reduce the sequence length of the acoustic features in the CTC compressor: ``blank-removal" and ``frame-averaging". For ``blank-removal", we simply discard all the frames that predicted the blank symbol according to the distribution of the CTC posteriors. On the other hand, for ``frame-averaging", we average the hidden states of consecutive frames without blank frames removed, once their CTC predictions belong to the same class. 

\textbf{Audio encoder}
The audio encoder is used to bridge representations generated from the CTC compressor to the text embeddings of the text-LLM. This module is designed to be relatively small in size and is initialized with random weights. During the fine-tuning process, the audio encoder is optimized to effectively integrate the audio information within the LLM, enhancing the overall performance of the system.
Different from the methods in ~\cite{huang2023audiogpt,rubenstein2023audiopalm}, where the audio encoder is trained to firstly map the speech signal into discrete tokens, which is then consumed by LLM, the proposed audio encoder is directly optimized to map the compressed acoustic signal to the continuous semantic space of LLM, allowing a deep integration between the audio encoder and the language model.


\textbf{Instruct learning}
For each training sample, we prepend a text prompt that briefly describes the task, e.g., ``$\mathtt{audio}$ $\Rightarrow$ $\mathtt{English}$'' and ``$\mathtt{transcribe}$ $\mathtt{the}$ $\mathtt{audio}$ $\mathtt{into}$ $\mathtt{English}$''. The text prompt are sampled from a pre-defined list, where some prompts contains the source language ID following the format ``$\mathtt{translate}$ $\mathtt{[source]}$ $\mathtt{audio}$ $\mathtt{into}$ $\mathtt{English}$''.
During evaluation, we fix the text prompt as ``$\mathtt{translate}$ $\mathtt{the}$ $\mathtt{audio}$ $\mathtt{into}$ $\mathtt{English}$'' for all testing samples.

\textbf{LoRA fine-tuning}
On top of the proposed model, we apply the LoRA to four attention matrices in each layer of the LLaMA Transformer  (e.g., $W_q, W_k, W_v, W_o$). To stabilize the training, we adopt a two-stage training scheme which means we train the audio encoder firstly with the CTC compressor and LLaMA frozen and then introduce LoRA to the well-trained model and perform the second stage optimization. The entire system is still trained with cross-entropy loss between the LLM output and the reference transcription sequence on the same training data.

\begin{figure}[!tbp]
\centering
\includegraphics[width=0.46 \textwidth]{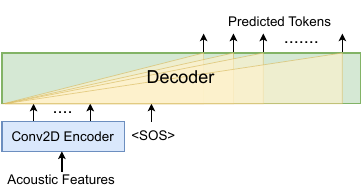}
\caption{The architecture of the decoder-only model for the from-scratch training. We use $\langle \text{SOS}\rangle$ token to indicate the starting of the text generation.}
\label{fig:res} 
\end{figure}
\textbf{From-scratch training}
To further explore the potential of decoder-only architecture as a foundational architecture for speech modeling, we also include a ``from-scratch" training of a decoder-only architecture. Here, we replace the text prompt, audio encoder, and CTC compressor with a randomly initialized convolutional 2D encoder. We also replace the pretrained LLaMA network with a much smaller randomly initialized autoregressive network. This architecture is shown in Figure 2. We add an $\langle \text{SOS}\rangle$ token at the end of the acoustic sequence to indicate the starting of the generation. In this case, the generation of the text sequence $\mathbf{y}$ with a decoder-only model is conditioned purely on audio signal $\mathbf{x}$ and previously generated text sequence $\mathbf{y}_{<n}$:
\begin{equation}
p(\mathbf{y}|\mathbf{x}; \Theta_\text{DEC}) = \prod_{n=0}^{N - 1} p(y_n|\mathbf{y}_{<n}, \mathbf{x}; \Theta_\text{DEC})
\end{equation}
where $\Theta_\text{DEC}$ refers to the parameters of the decoder model.

\label{ssec:subhead}

\section{Experiments}
\label{sec:typestyle}

The speech translation (ST) task \cite{vila2018end,sperber2020speech, xue2022weakly, Xue2022LargeScaleSE, llamasu}  has been chosen as the primary evaluation benchmark for assessing the proposed methods. In this task, the goal is to develop a system that can accurately translate spoken language from 13 source languages to English. 

\subsection{Data and metric}
\label{subsec: data_metric}

The 13 source languages we want to translate to EN are German (DE), Chinese (ZH), Arabic (AR), Spanish (ES), French (FR), Italian (IT),  Dutch (NL), Japanese (JA), Russian (RU), Portuguese (PT), Estonian (ET), Swedish (SV) and Slovenian (SL). We chose these languages based on availability of training and testing data. The training data for each language contains 1K hours of in-house speech data. To make the model more robust, we also include 1K hours of EN data, bringing the total to 14K hours. The original source transcriptions for non-English speech utterances are fed into an in-house translation service to generate the corresponding English transcriptions with both punctuation and capitalization. Those pseudo-label English transcriptions are used as the target transcription for ST task training. All the training data was anonymized with personally identifiable information removed.


Our speech translation models are evaluated on the 13 languages from the above list. The corresponding test sets are selected from CoVoST 2 dataset \cite{wang2020covost}. We evaluate the BLEU \cite{post-2018-call} scores for the performance comparison.

\subsection{Models configuration}
\label{ssec:model_details}
\subsubsection{CTC compressor}
 The CTC compressor contains 2 convolution-2D layers followed by 4 Transformer layers for a 4-times subsampling with 15.8M parameters in total. Each transformer layer has a 512-dimensional self-attention module with 8 heads and a 2048 dimensional feed-forward network (FFN). Each convolution 2D layer has a stride size of 2 and kernel size of 3. 
 We pre-trained CTC compressor with the paired speech and text data (i.e., ASR task) from 13 languages using the CTC objective function because that in our preliminary experiments, the BLEU score with ASR task training is much better than the one with the ST task training. Once CTC compressor is trained, the parameters are frozen during later training stages.


For comparison, we include a convolution-based subsampling module as a baseline, which shares the same architecture with the CTC compressor but with additional 3 1D convolution layers on top, allowing $4\times8=32$ times feature length reduction in total. The convolution-based subsampling is jointly trained with the audio encoder parameters. 



\subsubsection{Audio encoder}

The audio encoder consists of 4 Transformer layers, where each layer has the same setting as in the CTC compressor except that the output tensor of the last layer is converted to the dimension of 4096, in order to match the dimension of semantic embedding in LLaMA.


For each training sample, we concatenate the embeddings of the text prompt and representations from the audio encoder along the time axis and use that as the prefix feature sequence to feed to the LLaMA model to generate the target language (EN) transcriptions. 

Two attention mask strategies are explored within the LLaMA model. The first follows the language model training, where a causal, i.e., lower triangle attention mask is applied for each transformer layer to constrain the self-attention to not look into the future. 
As the proposed model is ``non-streaming" in nature, we also explore a non-causal full attention mask strategy for the prefix part only \cite{raffel2020exploring}, i.e., text prompt and audio encoder representations, to enable the full context learning on the acoustic information.


\subsubsection{LoRA fine-tuning}

We simply choose rank value of 2 for LoRA fine-tuning experiments according to the results of the LoRA work \cite{hu2021lora}, i.e.,  8 rank-2 matrices in the shape of $2 \times 4096$ are introduced to each LLaMA Transformer layer as an adaptor, which results in 2.1M more parameters in total.
The LoRA fine-tuning is conducted on a well-trained Speech-LLaMA model, where the CTC compressor and the LLaMA parameters are frozen. We still update the audio encoder to learn better representations together with the adapted LLaMA.

\begin{table*}[t]
\begin{center}
\centering
\caption{BLEU scores of the 13 languages on the baseline and the proposed models.}
\begin{tabular}{ c|c|c||c||c|c|c|c|c|c|c} 
\hline \hline
Model & \multicolumn{2}{c||}{Seq2seq}& Decoder-only& \multicolumn{6}{c}{Speech-LLaMA} \\
\hline
ID & B1 & B2 & D1 & E0 & E1 &  E2 & E3 & E4 & E5 & E6\\
\hline
Compressor & $-$ & $-$ & $-$ & $\times$ & \multicolumn{2}{c|}{CTC (remove)} & \multicolumn{4}{c}{CTC (average)} \\
\hline
Learnable \#Param. & \multicolumn{2}{c||}{240M} & 150M & 29M &  \multicolumn{2}{c|}{14M} & 14M & 16.1M & 14M & 16.1M \\
\hline
Prefix Non-causal Mask & $-$ & $-$ & $\checkmark$ & $\times$  & $\times$  & \checkmark & $\times$ &  $\times$ & \checkmark & \checkmark \\
\hline
LoRA & $-$ & $-$ &$-$ &$\times$ &  $\times$ &  $\times$ & $\times$ & E3 & $\times$ & E5 \\
\hline
LLaMA Rescore & $-$ & \checkmark & $-$ & $-$ &  $-$ & $-$ & $-$ & $-$ & $-$ & $-$ \\
\hline
 AR      & 22.8 & 24.9 &  21.2   &  16.9    & 24.6  & 24.7 &   24.6    &  26.3    &  25.9    &  28.2    \\ 
 DE      & 22.6 & 23.6 & 21.3& 16.9 & 22.6 & 22.8 & 24.3 & 26.0   & 25.4 & 27.1 \\ 
 ZH      & 7.0  & 7.2  & 6.7& 3.4  & 9.6  & 10.1 & 10.1 & 11.4 & 10.8 & 12.3 \\ 
 ES      & 23.7 & 24.9 & 22.7& 19.6 & 23.5 & 24.0 & 25.4 & 27.3 & 26.2 & 27.9 \\ 
 FR      & 21.8 & 22.7 & 20.6& 15.4 & 20.9 & 21.1 & 22.6 & 24.5 & 23.2 & 25.2 \\ 
 IT      & 20.7 & 21.6 & 19.8& 16.7 & 21.4 & 21.0 & 23.7 & 25.3 & 24.0 & 25.9   \\ 
 NL      & 34.6 & 36.0   & 35.2& 28.3 & 32.4 & 35.0 &  34.1 & 36.0   & 34.9 & 36.5 \\ 
 JA      & 15.3 & 15.7 & 16.3& 10.3 & 17.5 & 17.1 &  17.7 & 19.8 & 19.2 & 19.9 \\ 
 RU      & 26.4 & 27.7 & 26.0 & 22.8 & 31.0   & 32.0 & 33.3 & 35.5 & 34.3 & 36.8 \\ 
 PT      & 28.9 & 30.2 &27.2 & 22.8 & 26.8 & 27.7 & 29.2 & 31.3 & 30.2 & 32.0 \\ 
 ET      & 9.4  & 9.4  & 7.4& 15.4 & 17.0   & 18.3 & 17.2 & 18.1 & 18.0 & 18.7 \\ 
 SV      & 24.4 & 25.6 & 27.5& 26.3 & 25.3 & 28.8 & 26.7 & 27.4 & 27.2 & 29.0 \\ 
 SL      & 13.3 & 12.7 & 13.3& 22.2 & 20.3 & 22.9 & 22.8 & 22.2 & 22.1 & 22.7 \\ 
\hline
\bf{Average BLEU} $\uparrow$ & \bf{20.8} &  \bf{21.7} & \bf{20.4} & \bf{18.2} & \bf{22.5} & \bf{23.5} & \bf{24.0} & \bf{25.5} & \bf{24.7} & \bf{26.3}\\
\hline
\hline
\end{tabular}
\end{center}
\end{table*}

\subsubsection{Baseline}

\label{sssec:baseline}

We adopt a seq2seq \cite{Berard2016ST, weiss2017sequence} based speech translation model as a baseline. More specifically, we use the Whisper~\cite{radford2023robust} architecture with 240M parameters and train it on the 14K hour data mentioned in Section 4.1. It contains a 12-layer audio Transformer encoder and a 12-layer text Transformer encoder, where the attention dimension and head number is 768 and 12, respectively.  We optimize the model using cross-entropy as the primary objective function but also augment this architecture with a CTC loss on the encoder. We train the whole network end-to-end in a multi-task fashion. Please note that, for a fair comparison, we start the seq2seq training from scratch and do not initialize with pretrained open source Whisper weights. During beam search inference, we do a joint-decoding (prefix-decoding) \cite{hori-etal-2017-joint} with CTC. To make the comparison with LLaMA boot-strapped models more appropriate, we also present results with n-best rescoring of the seq2seq model with LLaMA. To accomplish that, we do a simple log-linear interpolation between the scores from seq2seq and LLaMA for each of the n-best hypotheses and then re-rank accordingly. We use $n=5$ for seq2seq beam-search decoding and the re-ranking experiments.


\subsubsection{From-scratch training}
In this setting, the structure of the convolutional 2D encoder contains 2 convolutional layers which is the same as the one in the CTC compressor, which introduces a 4-times subsampling rate. 
For the Transformer decoder, we follow the implementation of LLaMA, where pre-normalization, SwiGLU activation function \cite{shazeer2020glu} and rotary positional embeddings (RoPE) \cite{su2021roformer} are adopted. Similar to the configuration of the seq2seq baseline, each decoder layer contains a 12-head self-attention module with the 768 attention dimension. The dimension of the feed-forward network is set as 4076.

\subsection{Training and evaluation}
We extract an 80-dim log mel-filterbank using 25 msec window and 10 msec hop size as the acoustic features. Global mean and variance normalization is applied. All models were trained with AdamW optimizer \cite{loshchilov2017decoupled} with $\beta_1 = 0.9$ and $\beta_2 = 0.98$ on 16 V100 GPUs and a warmup and linear decay learning rate strategy is used. Batch size varies with the model size. CTC compressor was trained for 100K steps with source language transcriptions, tokenized by LLaMA's tokenizer. The peak learning rate was set to $0.001$. In the first stage training of Speech-LLaMA, We perform a 500K step training with a peak learning rate of $0.015$ while in the later LoRA fine-tuning stage, we use additional 100K optimization steps with a peak learning rate of $2e^{-4}$. 
The from-scratch decoder-only models were trained with a peak learning rate of $0.001$ for at most 300K steps. We use the beam search algorithm with a beam size 4 for the decoding of all the decoder-only models, unless noted otherwise. Both seq2seq and decoder-only models use English-only byte pair encoding (BPE \cite{kudo2018sentencepiece}) model for the tokenization which has a vocabulary size of 5,857 while the Speech-LLaMA models keep using LLaMA's tokenizer.


\section{Results and discussions}
The results of the experiments are presented in Table 1, where several observations can be gleaned.

\subsection{Baselines}

For baselines, we report results on 2 systems. B1 is a seq2seq model described in Section \ref{sssec:baseline} and B2 is B1 with LLaMA n-best rescoring. As expected \cite{li2023prompting}, a 0.9 better BLEU score can be observed from B2 system over B1. This suggests that shallow integration with LLM can still bring benefits to the speech models. 

\subsection{Deeper integration with LLaMA}
While shallow integration can boost performance, the gains using a deep integration technique like Speech-LLaMA should be much higher. Systems E1 $\sim$ E6 describe Speech-LLaMA models in various configurations. We can find all  Speech-LLaMA configurations significantly outperform the baselines with the limited learnable parameters, resulting in up to 4.6 absolute BLEU score improvement (21.2\% relative). These results show the efficacy of the proposed system and also suggests the necessity for deeper integration between the speech models and text-LLMs. 


\subsection{CTC compressor}
Results from system E0, E1 and E3 
describe the importance of CTC compressor for audio length reduction, in our design. Comparing E1 over E0, we obtain consistently better performance showing the effectiveness of  CTC compressor over the convolution one. This gain is despite the fact that CTC compressor is  frozen during the training while the convolution compressor was fine-tuned with the rest of audio encoder. 
One hypothesis for the better performance of CTC compressor is that it leverages the transcription of each source language during pre-training stage as we also observe that replacing the current CTC compressor model with the one trained with ST labels brings worse BLEU scores in our preliminary experiments. This observation also suggests that a potentially better performance might be obtained if the source transcription is also used during the training stage. We leave this line of exploration for future works. 

Within the CTC compressor, comparing system E3 over E1, the ``frame-averaging" strategy shows a 1.5 better average BLEU score over ``blank-removal" strategy. We believe that it is because the CTC compressor can't very reliably distill all relevant information into non-blank representations. Thus the frames selected by the CTC compressor might lose some acoustic information which cause the degradation of the performance. The averaging strategy is more robust to this compression error which aligns with the prior work \cite{gaido2021ctc}.

 \subsection{Effect of non-causal attention mask}
 It is expected that the full attention mask over text prompt and acoustic representations would usually result in better speech representation, and consequently better results. For each type of CTC compression strategy, our experiments demonstrate that using a non-causal attention mask over a causal mask can indeed bring gains. Comparing system E2 over E1, we see that switching to a non-causal mask brings an additional gain of 1.5 average BLEU score when using the ``blank-removal" strategy within CTC compressor. Similarly, comparing systems E5 over E3, we again observe a gain of 0.7 average BLEU score, when using the ``frame-averaging" strategy within CTC compressor. The gain with non-causal mask is understandably larger in ``blank-removal" strategy, since future acoustic information can help compensate for potential loss of information caused due to removal of frames corresponding to the blank symbol of the CTC loss. Even in LoRA fine-tuning systems, e.g., comparing E6 and E4, we can still observe a gain of 0.8 average BLEU score with non-causal mask applied.
 
\subsection{LoRA fine-tuning}
E4 and E6 represent our systems with LoRA fine-tuning. 
Comparing E4 over E3 shows the gains using LoRA fine-tuning when using a causal attention mask while comparing E6 over E5 show corresponding gains when using a non-causal attention mask. We can obtain an additional increase of 1.5 and 1.6 average BLEU score, respectively. Note that only 2.1M additional parameters are added as adaptors. Potentially better performance might be observed when larger rank is used. We leave this exploration for future works.

\subsection{Decoder-only vs Encoder-Decoder}

Finally, the results for the randomly initialized decoder-only model are shown as system D1 in Table 1. This model achieves only slightly worse (0.4 lower BLEU score) performance compared to the  seq2seq baseline. But the total parameter for the decoder-only model in our study is also significantly lower than the seq2seq baseline. We think that decoder-only architecture can be more parameter efficient than the encoder-decoder architecture. This is because a single module is used to learn representations for both source and target sequences in the former while separate modules (encoder \emph{and} decoder) are used to generate representations for source and target sequences in the latter. This sharing of parameters to process input and output jointly can bring out better parameter efficiency in the decoder-only architecture. 
Our results do seem to validate this theory. In future, we will conduct more extensive analysis of how model size effects performance in these 2 architectures. 


\section{Conclusion \& Future work}
In this work, we propose a method to infuse an off-the-shelf large language model with acoustic information. The proposed model presents a deep integration between the audio with the LLM by directly mapping acoustic representation into the semantic space of LLM. We also explore several practical aspects of the proposed model for better performance including compression of the acoustic feature, attention mask design and adapter fine-tuning. We show that on a 13 language to English speech translation task, the proposed model significantly outperforms a strong sequence-to-sequence baseline model. We also show that the decoder-only architecture, trained from scratch, can achieve comparable performance with around 40\% fewer parameters, which verifies the potential of decoder-only models for general speech-to-text modeling.
\label{sec:print}

\vfill

\bibliographystyle{IEEEbib}
\bibliography{strings,refs}

\end{document}